\documentclass[preprint,pre,showpacs,showkeys]{revtex4}%
\usepackage{amssymb}
\usepackage{amsmath}
\usepackage{amsfonts}

  \providecommand{\rbr}[1]{\left( #1 \right)}
\providecommand{\sqbr}[1]{\left[ #1 \right]}  \providecommand{\mt}[1]{\mathrm{#1}}
\providecommand{\mc}[1]{\mathcal{#1}} \providecommand{\tprod}[1]{\sideset{}{_{\otimes_{#1}}}\prod}

\def\ra{\rightarrow}

\begin{document}

\title[ ]{Skepsis on the scenario of Biological Evolution provided by stochastic models}
\author{Thomas Oikonomou}
\email{thoikonomou@chem.demokritos.gr}
\affiliation{Institute of Physical Chemistry, National Center for Scientific Research ``Demokritos", 15310 Athens, Greece\\ \\School of Medicine
Department of Biological Chemistry, University of Athens, Goudi, 11527 Athens, Greece}

\begin{abstract}
Stochastic models, based on random processes, may lead to power law distributions, which provide long range correlations. The observation of
power law behavior and the presence of long range correlations in biological systems has been demonstrated in various studies. The combination
of the two just mentioned results, theoretical and experimental, supports strongly the scenario of biological evolution across different
organisms. In the current Letter we explore in a general way, using the algebra of Nonextensive Statistics introduced by Tsallis and coworkers,
if the processes which are described by a class of stochastic models are really random and discuss the results with regard to a possible
biological evolution.
\end{abstract}
\eid{ }
\date{\today }
\startpage{1}
\endpage{1}
\maketitle

In the last years a variety of systems in diverse scientific fields has been explored presenting power law distributions like in biophysics
\cite{OikProvTirn08}, neurophysics \cite{NovikovEtAl97}, economics \cite{TsallisAnteneodoEtAl03}, turbulence
\cite{BeckLewisSwinney01,ArimArim02}, urban agglomerations \cite{MalacMendesLenzi02} and many others. Because of the ubiquity of power laws many
stochastic models have been studied \cite{Provata99,MesserAL05}, which are based on random processes and lead to the desired power law
distributions. These results have a great impact on our point of view about the rules that dominate in nature. Singularly, in biology the
success of deriving power laws through the consideration of random processes supports strongly the scenario of the biological evolution across
different species. These models do not give explicit description of the transition between different organisms during the evolution but they can
form well, under consideration of biological processes like mutation, duplication, etc., symbolic sequences which correspond to polynucleotide
DNA chains. For example in Ref. \cite{Provata99} Provata used a random aggregation model to describe a possible formation of such a
polynucleotide DNA chain. The model is based on the probability formula
%
\begin{align}\label{Takay}
P(s,t+\Delta t)&=\sum_{n=1}^{N}P(I)\binom{N}{n}\rbr{\frac{1}{N}}^n \rbr{1-\frac{1}{N}}^{N-n}\prod_{j=1}^{n}P(s_j,t)\Big|_{\sum_{j=1}^{n}s_j+I=s}
\end{align}
%
%
introduced by Takayasu and collaborators \cite{Takayasu1,Takayasu2}, which describes the probability of finding a macromolecule of size $s$ at
the time $t+\Delta t$. The analytical expression of Eq. (\ref{Takay}) is computed in the transformed Fourier space. A different stochastic
approach is considered in Ref. \cite{MesserAL05} by Messer, Arndt and L\"assig, who constructed a master equation for the joint probabilities
$P_{\mt{eq}}(r,t)$ and $P_{\mt{op}}(r,t)$ of finding a $GC$ pair and a $AT$ pair, respectively, at a distance $r$
%
\begin{align}\label{Messer}
\begin{aligned}
\frac{d}{dt}P_{\mt{eq}}(r,t)&=2\mu_{\mt{eff}}\sqbr{P_{\mt{op}}(r,t)-P_{\mt{eq}}(r,t)}\\
&+\sqbr{r\delta+(r-1)\gamma^{+}}\sqbr{P_{\mt{op}}(r-1,t)-P_{\mt{eq}}(r,t)}\\
&+r\gamma^{-}\sqbr{P_{\mt{eq}}(r+1,t)-P_{\mt{eq}}(r,t)}
\end{aligned}\quad.
\end{align}
%
The exchange $P_{\mt{eq}}(r,t)\leftrightarrow P_{\mt{op}}(r,t)$ in  Eq. (\ref{Messer}) gives the analogous equation for the probability
$P_{\mt{op}}(r,t)$. Then, the correlation function $C(r,t)=P_{\mt{eq}}(r,t)-P_{\mt{op}}(r,t)$ can be computed. The consideration of both Eqs.
(\ref{Takay}) and (\ref{Messer}) leads, under certain assumptions, to power law distributions.

On the other hand, it is well known that the existence of power law distributions indicates the presence of long (or short) range correlations.
Yet, this is in contradiction with the statistical assumptions of the stochastic models about random processes. It would mean that uncorrelated
processes can create, in an unexplained way, correlations between them. Even if we assume that in nature such an event (creation of correlations
through  random processes) can be possible, can we assume the same in mathematics? A stochastic model is a mathematical construction with
concrete rules and it can not lead to the existence of correlations if we have not first introduced them in the model. Starting from this point
of view we shall explore in the next paragraphs whether a class of stochastic models, where sums appear like in Eqs. (\ref{Takay}) and
(\ref{Messer}), describes really random processes. For this aim we shall use the $q$-Algebra \cite{Borges2004} of Nonextensive Statistics
\cite{BoonTsal05}. The results will be discussed in the frame of biology.

Nonextensive Statistics is a possible generalization of the ordinary random statistics, depending on a parameter $q\in[0,1]$ (the
$q_{\geqslant1}$-branche corresponds to different $q$-statistics \cite{Oik2007a}). Its structure is based on the following deformed logarithm
%
\begin{align}\label{GenLog1}
\ln_q(x)&:=\frac{x^{1-q}-1}{1-q},\qquad (x>0),
\end{align}
%
with its inverse deformed exponential function
%
\begin{align}\label{GenExp1}
\mt{e}_q(x)&:=\sqbr{1+(1-q)x}^{\frac{1}{1-q}},\qquad (x>0).
\end{align}
%
When $q=1$ the entire generalized statistical structure recovers the ordinary one. One of the main points in Nonextensive Statistics is its
successful treatment of correlated probabilities (or processes in general), which are projected on the values of the parameter $q$.

First, we introduce some elementary notions of the Probability Theory. A \emph{probability space}, which represents our uncertainty regarding an
experiment consists of two parts: i) a set of events/outcomes, which is denoted as \emph{sample space} $\Omega$ and ii) a real function of the
subsets of $\Omega$ which is denoted as \emph{probability measure} $P$. Now, let us follow the next thought. We consider two coins $I$ and $II$.
The sides of each coin have a different symbol, so that the sample spaces $\Omega_{I}$ and $\Omega_{II}$ consist of two events,
$\Omega_{I}=\{A,\,B\}$ and $\Omega_{II}=\{\tilde{A},\,\tilde{B}\}$. Then, the \emph{intersection probability} (of both occurring)  in each
sample space separately is of course equal to zero
%
\begin{align}\label{ProbTheory2}
P(A\cap B)&=0=P(\tilde{A}\cap \tilde{B}).
\end{align}
%
The \emph{union probability} (of either occurring)  in both spaces separately is equal to the unity
%
\begin{align}\label{ProbTheory1}
\begin{array}{c}
               P(A\cup B)=P(A)+P(B) \\
               P(\tilde{A}\cup \tilde{B})=P(\tilde{A})+P(\tilde{B}) \\
\end{array}\Bigg\}&=1.
\end{align}
%
Eq. (\ref{ProbTheory1}) corresponds in physics to the \emph{normalization constraint}. In this probability measure the event probabilities in
each sample space are strongly correlated in such a way that their sum is always equal to the unity. We denote these correlations as
correlations of Type I. On the other hand, with respect to the values of the probabilities, there are infinite ways to fulfil Eq.
(\ref{ProbTheory1}). In order to capture all these possibilities, or better to attribute them to a cause, we introduce yet another type of
correlations, which we call correlations of Type II. These correlations are individual for each event. In our example they could represent the
different weights of each side of the coin. When the correlations of Type II vanish, then we have equal event probabilities.

In further, we join the spaces $\Omega_{I,\,II}$. Now, we have one more type of correlations to consider, which we denote as correlations of
Type III. These correlations are between the several combinations of the events. Following these steps, the consideration of more complicate
processes or compositions would lead to an increasing number of the correlation types (higher order correlations, Type $>$III). In shake of the
simplicity, we assume that there are no correlations of the Types II and III in the new sample space $\Omega_{III}$ and because of this,
$\Omega_{III}$ is a tensor product of both subspaces $\Omega_{III}=\Omega_{I}\times\Omega_{II}$ $=$ $\{A\tilde{A},\,A\tilde{B},$
$B\tilde{A},\,B\tilde{B}\}$. In this case the event probabilities $P(A)$, $P(B)$, $P(\tilde{A})$ and $P(\tilde{B})$ are equal and independent.
Then, the intersection probabilities are given by
%
\begin{align}
\label{ProbTheory3a}
P(A\cap \tilde{A})&=P(A)P(\tilde{A}),\\
\label{ProbTheory3b}
P(A\cap \tilde{B})&=P(A)P(\tilde{B}),\\
\label{ProbTheory3c}
P(B\cap \tilde{A})&=P(B)P(\tilde{A}),\\
\label{ProbTheory3d} P(B\cap \tilde{B})&=P(B)P(\tilde{B}),
\end{align}
%
and because of the union probability
%
\begin{align}\label{ProbTheory4}
\begin{aligned}
P(A\tilde{A}\cup A\tilde{B}\cup B\tilde{A}\cup B\tilde{B})&=P(A)P(\tilde{A})+ P(A)P(\tilde{B})+\\
&\quad \, P(B)P(\tilde{A})+P(B)P(\tilde{B})=1,
\end{aligned}
\end{align}
%
their values are equal to 1/4. Here, there is a very important point to stress. Namely, the feature of the probability (in)dependence is
meaningful in the frame of an appropriate probability measure. As we can see in Eqs. (\ref{ProbTheory3a}) - (\ref{ProbTheory3d}), the
intersection probabilities of $\Omega_{III}$ are computed based on the independence of the event probabilities between the two subspaces
$\Omega_{I}$ and $\Omega_{II}$. Yet, the union probability of $\Omega_{III}$ correlates strongly the intersection probabilities. Furthermore, it
becomes evident that when we refer to correlations we have to be very concrete about their Type. To give a more general character to the
intersection probabilities for the further purposes of this Letter we shall call them \emph{Blocks of probabilities}.

In Ref. \cite{Oik2007a} the present author demonstrated the importance of the distinction between the diverse types of correlations. The
existence of two different types of correlations within Nonextensive Statistics  has been introduced, the \emph{inner} and the \emph{outer},
described by a set of parameters $\mc{R}=\{R_i\}_{i=1,\ldots,n}$ and $\mc{Q}=\{Q_i\}_{i=1,\ldots,n'}$ respectively, in order to distinguish
between the three generalized entropic structures that can be created, based on any generalized logarithmic function. The inner and outer
correlations in this case should not be necessarily identified with ones in our example above. Their notion must be every time  adjusted to the
features of a problem under consideration. For the one parametric ($\mc{Q}=\{Q_1\}=q$) generalized logarithm (\ref{GenLog1}) with its inverse
function (\ref{GenExp1}) these three structures correspond to the Tsallis, R\'enyi and Nonextensive Gaussian entropy. We shall keep the notation
of the correlation-sets $\mc{R}$ and $\mc{Q}$ throughout the paper.

For an arbitrary deformed logarithmic and exponential function we define the generalized product between two elements $x,y\in\mathbb{R}_{+}$ as
\cite{Oik2007a}
%
\begin{align}\label{QProd0}
x\otimes_{\mc{Q}}y&:=\mt{e}_{\mc{Q}}\rbr{\ln_{\mc{Q}}(x)+\ln_{\mc{Q}}(y)}.
\end{align}
%
For $\mc{Q}\ra\mc{Q}_0$ we obtain the ordinary multiplication since $\mt{e}_{\mc{Q}_0}\rbr{\ln_{\mc{Q}_0}(x)+\ln_{\mc{Q}_0}(y)}$
$=\mt{e}\rbr{\ln(x)+\ln(y)}$ $=x\times y$. Then, for the $\mc{Q}$-multiplication of $m$ elements we have
%
\begin{align}\label{QProd}
\tprod{\mc{Q}}_{k=1}^{m}x_{k}&=\mt{e}_{\mc{Q}}\rbr{\sum_{k=1}^{m}\ln_{\mc{Q}}(x_k)},\qquad (x_k>0).
\end{align}
%
Using the deformed functions in Eqs. (\ref{GenLog1}) and (\ref{GenExp1}), Eq. (\ref{QProd}) takes the explicit form
%
\begin{align}\label{QProd2}
\tprod{q}_{k=1}^{m}x_{k}&=\sqbr{\sum_{k=1}^{m}x_k^{1-q}-(m-1)}^{\frac{1}{1-q}}.
\end{align}
%
If we express the normalization constraint of $W$ intersection probabilities $P_j$ through the $q$-algebra we can easily verify the relation
%
\begin{align}\label{NormCon}
\sum_{j=1}^{W}P_j&=W\otimes_{q}\tprod{q}_{j=1}^{W}P_j\Bigg|_{q=0}=1.
\end{align}
%
The value $q=0$ in $q$-Statistics corresponds to long range correlations. This result reproduce correctly the correlations of Type I behind the
normalization constraint, which relate all probabilities in a very specific way. Since generalized statistics gives us proper results and does
not lead to inconsistencies we shall use it in the next paragraph in order to complete the aim of this Letter.

We consider the statistical processes of a system and define the probabilities under which these processes take place and the way they interact
(correlated or uncorrelated). An example for such a system under consideration could be a DNA chain and a possible statistical process could be
the mutation of a nucleotide. In connection to our example with the coins, all these interacting probabilities define the respective Blocks of
probabilities for the current approach and are denoted as $\{A^{(\mc{R})}_{k}(P_j)\}_{k=1,\ldots,m}$. The internal $\mc{R}$-correlations
characterize the different types of correlations within a Block. These correlations are canceled for certain values $\mc{R}\ra\mc{R}_0$. Now, we
want to see how the several $k$-Blocks interact with each other. These interactions are described by the external correlations $\mc{Q}$. Then,
we introduce the probability functional $B_{\{\mc{Q},\,\mc{R}\}}(P_j)$ which is equal to the entire internal and external interactions. It could
be a derivative of the probabilities $P_j$ in $A^{(\mc{R})}_k(P_j)$, as in Eq. (\ref{Messer}), or simply a new probability, as in Eq.
(\ref{Takay}), etc. The elimination of the external correlations is defined again in a certain limit $\mc{Q}\ra\mc{Q}_0$. Then, the entire
process is given by an expression of the form
%
\begin{align}\label{InTest1}
B_{\{\mc{Q},\,\mc{R}\}}(P_j)&=m\otimes_{\mc{Q}}\tprod{\mc{Q}}_{k=1}^{m}A^{(\mc{R})}_{k}(P_j),
\end{align}
%
with $B_{\{\mc{Q},\,\mc{R}\}}(P_j)>0$. We stress that the Eq. (\ref{InTest1}) does not describe exact processes. It is a general expression of a
set of probability interactions. The $\mc{Q}$-multiplication with the constant $m$, does not have a specific meaning but is useful for our
purpose. We demand that the stochastic processes inside of each Block are independent ($\mc{R}\ra\mc{R}_0$), so that they are multiplicatively
connected through the ordinary multiplication. Using Eq. (\ref{QProd2}), the relation in Eq. (\ref{InTest1}) can be written as
%
\begin{align}\label{InTest2}
B_{\{q,\,\mc{R}_0\}}(P_j)&=\Bigg[m(q)+\sum_{k=1}^{m}\sqbr{A^{(\mc{R}_0)}_k(P_j)}^{1-q}\Bigg]^{\frac{1}{1-q}}.
\end{align}
%
with $m(q):=m^{1-q}(1-m^q)$. In further, we set $q=0$ as we have done in Eq. (\ref{NormCon}). Then, from Eq. (\ref{InTest2}) we obtain
%
\begin{align}\label{InTest3}
B_{\{q=0,\,\mc{R}_0\}}(P_j)&=\sum_{k=1}^{m}\sqbr{A^{(\mc{R}_0)}_k(P_j)}.
\end{align}
%
Eq. (\ref{InTest3}) is our main result. Namely, when the probability measure $B_{\{q,\,\mc{R}_0\}}(P_j)$ is equal to the sum ($q=0$) of the
probability Blocks, even if the internal processes inside a Block are independent, then this measure describes long range correlations.
Accordingly, every stochastic model, which is based on such a sum, gives a possible creation of sequences through a very concrete dynamic, where
the processes of the several probability Blocks are long range correlated (synchronized). Comparing Eqs. (\ref{Takay}) and (\ref{Messer}) with
Eq. (\ref{InTest3}), we see that the two earlier equations are specific expressions of the latter one.

If we accept that biological evolution, as described by Eq. (\ref{InTest3}), has taken place, then the transition between different species
during the evolution is attributed to the combination of two different kinds of modifications in the $\mc{R}$-correlations in each Block. The
first kind of modifications do not affect the structure of the given dynamic. This case would correspond to the creation of different organisms
of the same class. The second kind of modifications does affect the structure of the given dynamic. This case would correspond to the creation
of different classes of organisms. However, the existence of long range correlations has the effect that a modification of a single probability
causes simultaneously changes in all involved terms. This has as consequence that every new created organism does not have the time to be
adapted to its environment, so that it is able to survive. On the other hand, the adaption to the environment is the basis of the natural
selection and the biological evolution. As we can see, the introduction of long range correlations gives a problematic to the scenario of
biological evolution, as has been assumed up to now.

Summarizing, we have shown that the dependence or independence between probabilities is associated to an appropriate probability measure. The
sum of probabilities or of a Block of probabilities corresponds to long range interactions between the participating terms. Accordingly, any
stochastic model in which such a sum appears, correlates in a very specific way, with long range features, the respective processes. The result
is a non-exponential probability distribution function. Beside the clarification of the appropriate probability measure,  we need to clarify the
Type (or Types) of the correlations as well. There is a variety of different types of correlations in a process, whose confusion may lead to
incorrect results. The more complicated a process is, the more different types of correlations are involved in this process. The presence of
correlations creates difficulties considering the cornerstone of the natural selection in biology, which is the adjustment of an organism to its
environment.


\end{document}